\documentclass{PoS}

\title{
\vspace*{-2cm}
\begin{minipage}{\textwidth}
\begin{flushright}
\texttt{\footnotesize
PoS(LAT2007)237\\%
CERN-PH-TH/2007-121\\%
}
\end{flushright}
\end{minipage}\\[15pt]
\vspace*{+2cm}
\mbox{On the phase diagram of QCD at finite isospin density}}

\ShortTitle{On the phase diagram of QCD at finite isospin density \hfill}

\author{Philippe de Forcrand\\
  Institute for Theoretical Physics, ETH Zurich, CH--8093 Zurich, Switzerland \\
  and \\
  CERN, Physics Department, TH Unit, CH--1211 Geneva 23, Switzerland \\
  E-mail: \email{forcrand@phys.ethz.ch}}

\author{Mikhail A.~Stephanov\\
  Physics Department, University of Illinois, 845 W. Taylor St., Chicago, IL 60607-7059, USA\\
  E-mail: \email{misha@uic.edu}}

\author{\speaker{Urs Wenger}
  \\
  Institute for Theoretical Physics, ETH Zurich, CH--8093 Zurich,
  Switzerland\\
  E-mail: \email{wenger@phys.ethz.ch}}

\abstract{Using a canonical formalism, we determine the equation of state and
  the phase diagram of eight-flavour QCD, as a function of temperature and
  isospin density. Two mechanisms are at work: Bose condensation of pions at
  high density, and deconfinement at high temperature. We study their
  interplay and find that on our small and coarse lattice the first order
  deconfinement transition appears to end at a critical point at finite
  density. We investigate the strength of the overlap and of the sign
  problems and discuss implications for the baryonic density case.}

\FullConference{The XXV International Symposium on Lattice Field Theory\\
		 July 30 - August 4 2007\\
		 Regensburg, Germany}

\begin{document}
%%%%%%%%%%%%%%%%%%%%%%%%%%%%%%%%%%%%%%%%%%%%%%%%%%%%%%%%%%%%%%%%%%%%%%%%%%%%%
\section{Introduction}
QCD at finite isospin density is interesting for several reasons. First of all
it is a special case of the physically relevant situation where $\mu_u \neq
\mu_d \neq \mu_s$. Secondly, it serves as a platform to assess indirectly the
limitations of the various numerical approaches to finite {\it baryon}
density. And thirdly it provides a rich and exciting range of physical
phenomena: at low temperature and density there is a pion gas which transforms
into a quark gas at high temperature via a deconfinement phase transition,
while at low temperature and high density there is Bose condensation of
charged pions.

The transition at finite isospin density is, just like in the case of finite
baryonic density, due to a large density of a conserved charge, namely the
isospin charge $Q$. However, the system does not carry baryon number, but
instead the chemical potentials of quarks $u$ and $d$ are set equal in
magnitude, but opposite in sign, i.e.~$\mu_u = -\mu_d = \mu_I/2$.
Furthermore, taking the $u$ and $d$ quarks masses equal, the system is
accessible through lattice simulations because the positivity of the theory is
guaranteed by
\begin{equation}
\label{eq:positivity}
\tau_1 \gamma_5 D\ \gamma_5 \tau_1 = D^\dagger.
\end{equation}

Using this positivity and QCD inequalities, Ref.~\cite{Son:2000xc} showed that
the symmetry breaking related to the density driven transition must be
accompanied by a condensate $\langle \bar \psi i \gamma_5 \tau_{1,2} \psi
\rangle$, i.e.~by $\pi^- \sim \bar u \gamma_5 d,\ \pi^+ \sim \bar d \gamma_5
u\ $ states. On the lattice, however, this interesting physical system has
largely been ignored, with the notable exception of a series of investigations
by Kogut and Sinclair \cite{Kogut:2004zg,Sinclair:2006zm,Sinclair:2007ce}.

%%%%%%%%%%%%%%%%%%%%%%%%%%%%%%%%%%%%%%%%%%%%%%%%%%%%%%%%%%%%%%%%%%%%%%%%%%%%%
\section{Expectations}
At small isospin densities one can use chiral perturbation theory to describe
the physics of the system. The chiral Lagrangian is given by
\begin{equation}
\label{eq:eft_lagrangian}
{\cal L} = \frac{1}{4} f_\pi^2 \textrm{Tr}[{\cal D}_\mu\Sigma {\cal D}_\mu\Sigma
^\dagger - 2 m_\pi^2 \textrm{Re} \Sigma]
\end{equation}
where $\Sigma \in \textrm{SU}(2)$ is the matrix pion field. The isospin
chemical potential $\mu_I \tau_3$ can be included in ${\cal D}_0$ at leading
order without introducing additional low energy constants and breaks
$\textrm{SU}(2)_{L+R} \rightarrow \textrm{U}(1)_{L+R}$.  The effective
potential from the Lagrangian above can be minimised as a function of $\mu_I$
using
\begin{equation}
% \label{eq:condensate}
\overline 
\Sigma = \cos \alpha + i (\tau_1 \cos \phi  + \tau_2 \sin \phi) \sin
\alpha 
\end{equation}
where $\phi$ is an irrelevant flavour rotation angle corresponding to the
residual $\textrm{U}(1)_{L+R}$ symmetry and $\alpha$ parametrises the rotation
from the standard chiral condensate $\langle \bar u u + \bar d d\rangle = 2
\langle \bar \psi \psi\rangle_0$ into the pion condensate $\langle \bar d
\gamma_5 u \rangle \neq 0$.

From the effective field theory description we can identify two fundamentally
different regimes. For $|\mu_I| < m_\pi$ no pions can be excited from the
vacuum and we have $\overline \Sigma = 1$, i.e.~$\langle \bar u u + \bar d
d\rangle = 2 \langle \bar \psi \psi\rangle_0$. This regime corresponds to the
standard QCD vacuum at low temperature and zero density.

For $\mu_I \ge m_\pi$, $\pi^+$ particles can be excited from the vacuum and a
Bose condensate of $\pi^+$ forms where $\langle \bar d \gamma_5 u \rangle \neq
0$. In that regime the remaining $\textrm{U}(1)_{L+R}$ symmetry is broken
spontaneously and hence the transition belongs to the universality class of
the $3d$ $XY$-model if it is second order.

Finally, for $\mu_I > m_\rho$ the chiral expansion breaks down and the
effective field theory description eq.~(\ref{eq:eft_lagrangian}) is no longer
applicable. Nevertheless, on general ground one can argue \cite{Son:2000xc}
that the condensate remains $\langle \bar d \gamma_5 u \rangle \neq 0$ and one
might expect a BEC-BCS crossover at large $\mu_I$ when the pions "dissociate"
(the QCD running coupling becomes weak).

A complete description of the various phases is provided by the equation of
state (EoS) of the theory. Here we do not attempt to calculate the full EoS
but rather the relation between the isospin density and the chemical
potential\footnote{In a slight misuse of notation we will nevertheless refer
  to eq.~(\ref{eq:EoS}) as EoS in the following.},
\begin{equation}
\label{eq:EoS}
\rho_I \equiv \frac{Q}{V} = \rho_I(\mu_I) .
\end{equation}
The canonical free energy $F(Q) = - \ln Z_C(Q)$ and its derivative
\begin{equation}
\label{eq:free_energy_difference}
F(Q) - F(Q-1) \quad \stackrel{V\rightarrow \infty}{\Longrightarrow} \quad \frac{d F}{ d\rho_I } = \mu_I
\end{equation}
are numerically well accessible from a canonical formulation of lattice QCD.

\begin{figure}[t]
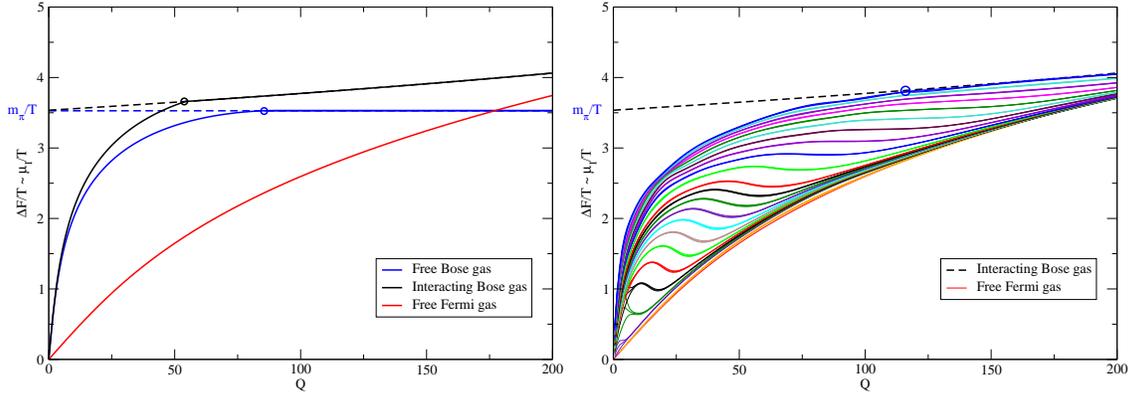

\begin{center}
\includegraphics[width=0.49\textwidth,angle=0]{Figures/dF_vs_Q_freeFermiGas.eps}
\includegraphics[width=0.49\textwidth,angle=0]{Figures/dF_vs_Q_Nf8_QCD_12.eps}

\end{center}
\caption{The free energy difference $(F(Q)-F(Q-1))/T$ versus the number of
  $u$-quarks $Q$.  
  {\it Left:} for a gas of free (blue) and weakly interacting
  (black) pions and for a gas of free massless fermions (red).  
  {\it Right:}
  simulation results on $8^3 \times 4$ lattices for successive temperatures,
  increasing from $\sim \frac{T_c}{2}$ to $\sim 1.1 T_c$ from top to bottom. }
\label{fig:dF_vs_Q}
\end{figure}

The behaviour of the system in the various expected phases can be anticipated
using simple semiquantitative considerations. For example, in the confined
regime at low temperature and low density, we have a gas of (to a first
approximation) free pions.  For bosons of mass $m= \hat m T$ and a chemical
potential $\mu = \hat \mu T$, the density reads
\begin{equation}
\rho_\textrm{free}(\hat{\mu},\hat{m})  = 
\frac{T^3}{2\pi^2} \int_0^{+\infty}
\hspace{-0.5cm} d\hat{p} \, \hat{p}^2 \left(
  \frac{1}{e^{(\omega-\hat{\mu})} - 1} -
  \frac{1}{e^{(\omega +\hat{\mu})} - 1} \right) 
\end{equation}
where $\omega = \sqrt{\hat{p}^2+\hat{m}^2}$. The behaviour at zero and low
temperature is sketched in Fig.~\ref{fig:dF_vs_Q} where we plot the free
energy difference, eq.~(\ref{eq:free_energy_difference}), versus the number of
$u$-quarks $Q$. At $\mu_I = m_\pi$ the system Bose condenses and as a
consequence the free energy difference remains constant beyond that point
(denoted by the circle in Fig.~\ref{fig:dF_vs_Q}).
For an interacting gas of pions the 'EoS' can be described by the effective
field theory eq.~(\ref{eq:eft_lagrangian}) and the density at zero temperature
is given by
\begin{equation}
\label{eq:pion condensate}
\rho_I = f_\pi^2 \mu_I \left(1 - (\frac{m_\pi}{\mu_I})^4\right).
\end{equation}
The interaction leads to an increase of the free energy difference at a given
density and temperature as sketched in Fig.~\ref{fig:dF_vs_Q}.

In the high temperature deconfined phase and for quark mass $m_q \ll \mu_I$
the pressure can be approximated by that of a massless, free Fermi gas,
leading to
\begin{equation}
\label{eq:free_fermi_gas}
\frac{\rho_I}{T^3} = \hat \mu_I + \frac{1}{\pi^2}~
\hat \mu_I{}^3 \, .
\end{equation}
The corresponding situation is sketched in Fig.~\ref{fig:dF_vs_Q}.

%%%%%%%%%%%%%%%%%%%%%%%%%%%%%%%%%%%%%%%%%%%%%%%%%%%%%%%%%%%%%%%%%%%%%%%%%%%%%
\section{Numerical results}
For computational convenience we consider in our numerical simulations two
staggered fermion fields corresponding to $N_f=8$ QCD in the continuum,
i.e.~four $u$- and four $d-$quark species, all degenerate in mass. We keep
the lattice volume fixed to $8^3 \times 4$ as well as the bare quark mass $a m
= 0.14$ while varying the bare coupling $\beta$ and the isospin chemical
potential $a \mu_I$. Note that the bare quark mass is tuned in such a way that
the deconfinement transition at $T_c$ and zero density is first order. Varying
$\beta$ changes the lattice spacing $a$ and thereby the temperature $T =
\frac{1}{4a}$ and we cover a range of temperatures between $\frac{1}{2} T_c
\lesssim T \lesssim T_c$. We emphasise that for the range of couplings we
consider, the pion mass $a m_\pi$, measured at zero temperature, changes very
little in units of the lattice spacing, so that the ratio $m_\pi/T$ stays
almost constant.  Note that, contrary to
Refs.~\cite{Kogut:2004zg,Sinclair:2006zm,Sinclair:2007ce}, we do not introduce
a $U(1)$-breaking source term in our action. Thus, we can observe the
spontaneous symmetry breaking characteristic of Bose condensation without
performing any delicate extrapolation.  Finally, our results are obtained by
combining 68 ensembles at six values of $\mu_I$ up to $\mu_I/T \lesssim 5$
with Ferrenberg-Swendsen reweighting.

The free energy is obtained from the canonical partition function, $F(Q) =
-1/T \log Z_C(Q)$. $Z_C(Q)$ is estimated using standard grand canonical Monte
Carlo simulations
\begin{equation}
\frac{Z_C(Q)}{Z_{GC}(\mu_I)} = \left\langle \frac{|\hat{\det_Q}|^2}{|\det
  D(\mu_I)|^2} \right\rangle_{Z_{GC}} \; 
\end{equation}
where $\hat{\det_Q}$ is obtained by decomposing exactly the part of the
measure which depends on $\mu_I$ \cite{Hasenfratz:1991ax,deForcrand:2006ec}, i.e., the fermion determinant,
\begin{equation}
\det D(\mu_I) = \sum_{Q=-3V}^{+3V} \hat{\det{}_Q} ~ \exp(Q \mu_I/T) \; .
\end{equation}
Our results for the free energy difference
eq.~(\ref{eq:free_energy_difference}) (in units of $T$) versus the number of
$u$-quarks $Q$ for successive temperatures together with the theoretical
expectations are shown in Fig.~\ref{fig:dF_vs_Q} right. At high temperature
(lowest curve) our data is very well described by an ansatz motivated by the
free Fermi gas description, eq.~(\ref{eq:free_fermi_gas}), where the
coefficients for the linear and cubic term are multiplied by factors that take
into account effects of interactions. The numerical values we obtain are
reasonably close to one and consistent with those obtained in
\cite{Kratochvila:2006jx}.

\begin{figure}[t]
\begin{center}
\includegraphics[width=0.80\textwidth,angle=0]{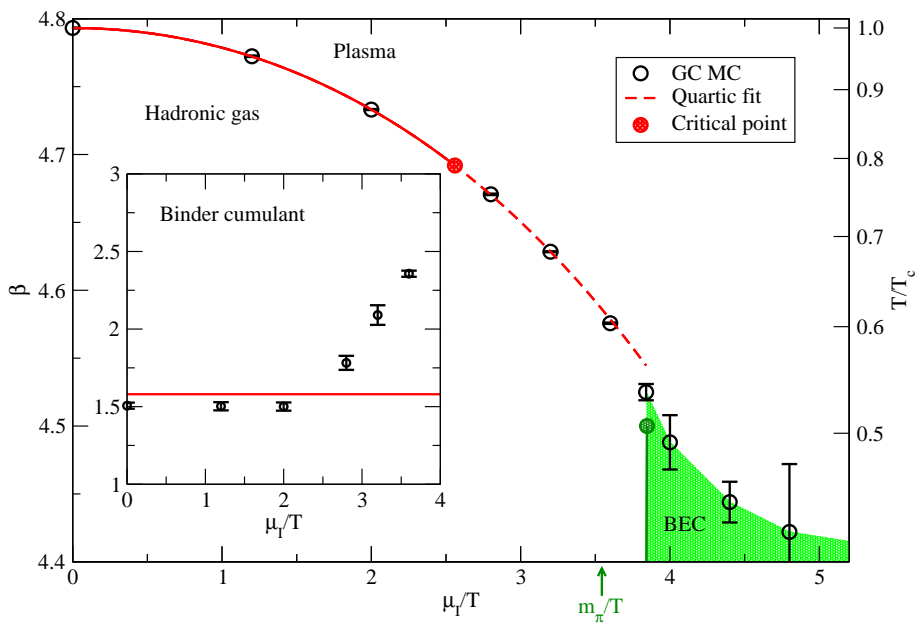}
\end{center}
\caption{Phase diagram in the $(\mu_I/T,T/T_c)$-plane from grand-canonical
  Monte-Carlo simulations (GCMC) at finite isospin chemical potential.  The
  red circle indicates the critical point as inferred from the change of the
  behaviour of the Binder cumulants (shown in the inset for the quark
  condensate) at $\mu_I/T \simeq 2.5$. The green circle indicates the
  transition into the pion condensed phase (BEC) as obtained from the
  canonical analysis.}
\label{fig:phase_diagram_B4_2}
\end{figure}

Decreasing the temperature successively, we obtain the second-, third-, ...
to-lowest curves. The S-shape of the curves for $T \leq T_c$ indicates that
$\rho_I(\mu_I)$ is multivalued, and is thus characteristic of a first order
transition -- in the thermodynamic limit the density would jump at the
critical isospin chemical potential. Here, the critical $\mu_I$ can be
determined using the Maxwell construction and it increases as the temperature
is decreased. Furthermore, the S-shape becomes smoother and finally disappears
at around $T/T_c \simeq 0.8$ where the jump in the density vanishes and the
first order transition is replaced by a crossover. In order to ascertain the
presence of this critical point, we also monitor the distributions of the
plaquette, the Polyakov loop, the quark condensate and density by measuring
the corresponding Binder cumulants in grand-canonical simulations at fixed
$\mu_I$. These cumulants remain essentially constant in the region $0 \leq
\mu_I/T \leq 2$ beyond which they start to grow linearly, cf.~inset in
Fig.~\ref{fig:phase_diagram_B4_2}.  Attributing this behaviour to the critical
point of the 3d Ising model universality class one can locate the point at
$\mu_I/T \simeq 2.5$.  A finite-size scaling analysis and a continuum
extrapolation are of course required to locate this critical point reliably,
in particular, to disentangle it from the 2nd order BEC transition.  Our
findings are similar to those of Ref.~\cite{Sinclair:2006zm}, which for
$N_f=3$ also find that the Binder cumulant grows as an isospin chemical
potential is on.

Lowering the temperature further, we observe a transition from the pion gas
into the Bose condensed phase, marked by a circle in
Fig.~\ref{fig:dF_vs_Q} (right). It turns out that the system can very well be
described by the ansatz in eq.~(\ref{eq:pion condensate}) where the value of
$m_\pi$ is fixed from our measurement at zero temperature and $f_\pi$ is
treated as a free parameter.

The full phase diagram in the $(\mu_I/T,T/T_c)$-plane is presented in
Fig.~\ref{fig:phase_diagram_B4_2} where the black circles denote our results
from the grand-canonical Monte-Carlo simulations (GCMC) and the red circle
marks the critical point. For $\mu_I \leq m_\pi$ the pseudo-critical line
between the hadronic gas and the plasma can be well described by a quartic fit
(dashed line).  The green point marks the entry into the pion condensation
phase obtained from Fig.~\ref{fig:dF_vs_Q} (right). As expected, for $T > 0$
this point is shifted towards larger values of $\mu_I$ as compared to the zero
temperature situation where the transition would occur exactly at $\mu_I/T =
m_\pi/T$ (indicated by the small arrow on the $x$-axis).

\begin{figure}
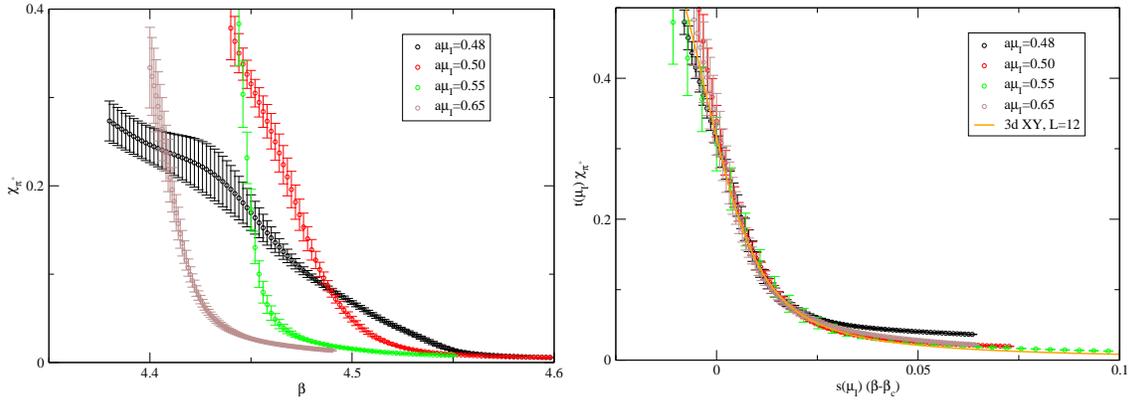

\begin{center}
\includegraphics[width=0.49\textwidth,angle=0]{Figures/chi_Piplus_unscaled.eps}
\includegraphics[width=0.49\textwidth,angle=0]{Figures/chi_Piplus_scaled.eps}
\end{center}
\caption{{\it Left:} Pion susceptibility $\chi_{\pi^+}$ for various values of
  $\mu_I$. {\it Right:} Data collapse obtained from shifting $\beta$ to
  $\beta_c$ and rescaling both the $x-$ and $y-$axis. Also shown is the
  universal function from simulations of the $3d$ $XY$-model.}
\label{fig:chi_Piminus_scaled_xy}
\end{figure}

The transition from the plasma to the pion condensation phase can be
determined by monitoring the pion susceptibility $\chi_{\pi^+} = \sum_x
\langle \pi^+(0)\pi^+(x) \rangle$. Since this transition is characterised by
the breaking of a $\textrm{U}(1)$ symmetry, it is expected to belong to the
universality class of the $3d$ $XY$-model.  We can expose the universal
behaviour of the transition by comparing the data at different values of
$\mu_I$ with the universal scaling curve obtained from simulations of the 3d
$XY$-model. Fig.~\ref{fig:chi_Piminus_scaled_xy} shows that the agreement is
excellent.

Finally, our calculation also provides a crosscheck of the reweighting method
from zero to finite (isospin or baryonic) chemical potential. It turns out
that for isospin chemical potential reweighting from $\mu_I=0$ is possible,
but fails to describe the phase diagram reliably. The overlap of the $\mu_I=0$
ensembles with the ones relevant at finite density is not sufficient to
determine e.g.~the order of the phase transition at finite $\mu_I$, and the
extrapolation to finite $\mu_I$ noticeably underestimates the density of the
system. This is understandable from the fact that one tries to extract finite
density information from the tail of a density distribution which is generated
at $\mu_I = 0$, i.e.~centered at zero.

\begin{figure}
\begin{center}
\includegraphics[height=5cm,angle=0]{Figures/dF_vs_Q_Nf8_QCD_mu0.00.eps}
\hspace{0.85cm} \includegraphics[height=5cm]{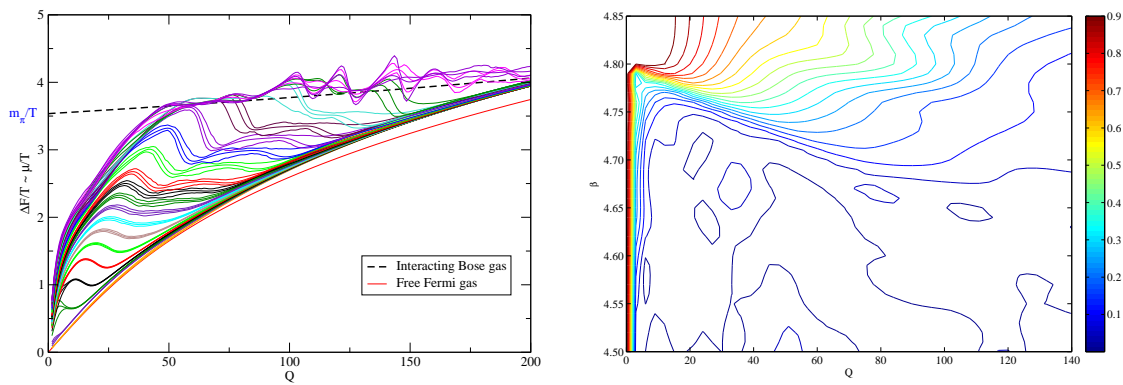}
\end{center}
\caption{{\it Left:} same as Fig.~1 right, but reweighted from zero density
  simulations only. {\it Right:} isocontour lines of the average sign $\langle
  \hat{\det_Q}^2 / |\hat{\det_Q} |^2\rangle$.}
\label{fig:sign}
\end{figure}

For an extrapolation to finite baryonic density one has the additional problem
that the average sign of the fermionic determinant tends to zero very quickly
with $\mu$, as illustrated in Fig.~\ref{fig:sign}, hence narrowing the range
of applicability of the reweighting procedure even further.
 
%%%%%%%%%%%%%%%%%%%%%%%%%%%%%%%%%%%%%%%%%%%%%%%%%%%%%%%%%%%%%%%%%%%%%%%%%%%%%
\section{Conclusions}
We determined the EoS and the phase diagram of $N_f=4+4$ QCD at finite isospin
density and finite temperature. The two mechanisms at work are clearly
exposed: we observe Bose condensation of pions at high density and
deconfinement at high temperature. For our quark mass, the transition from the
hadronic gas to the quark-gluon plasma is first order at zero density. As the
density is increased, the transition appears to turn into a crossover at
$\mu_I/T \simeq 2.5$, in qualitative similarity with the results of
Ref.~\cite{Sinclair:2006zm}. We caution, however, that to confirm these
findings a finite-size scaling analysis is still needed.

The transition from the hadronic gas to the BEC phase is determined from a
canonical analysis of the free energy density, occurs at low temperature
$T/T_c \lesssim 0.5$ at $\mu_I$ slightly larger than $m_\pi$ and is consistent
with being second order. The transition from the BEC phase to the plasma phase
is also second order and its universal behaviour is confirmed to be that of
the $3d$ $XY$-model.

Finally, by comparing our (interpolated) results for the EoS $\rho_I(\mu_I)$
with the ones obtained by reweighting from zero to finite $\mu_I$, we find
that the latter technique systematically underestimates the densities as the
chemical potential grows, and that to determine its range of reliability a
comparison of different approaches is needed.

%%%%%%%%%%%%%%%%%%%%%%%%%%%%%%%%%%%%%%%%%%%%%%%%%%%%%%%%%%%%%%%%%%%%%%%%%%%%%
\section*{Acknowledgements} 
We thank Joe Kapusta, Kasper Peeters, Krishna Rajagopal and Dam Son for
discussions, the Center for Theoretical Physics, MIT, and the Isaac Newton
Institute, Cambridge, for hospitality and the Minnesota Supercomputer
Institute for computer resources. M.A.S.~is supported, in part, by DOE grant
No.\ DE-FG02-01ER41195.

%%%%%%%%%%%%%%%%%%%%%%%%%%%%%%%%%%%%%%%%%%%%%%%%%%%%%%%%%%%%%%%%%%%%%%%%%%%%%
\bibliographystyle{plain}
\bibliography{pdQCDfid}

\end{document}